# Sweeping Plasma Frequency of Terahertz Surface Plasmon Polaritons with Graphene

*Mingming Feng[1][a], Baoqing Zhang[1][a], Haotian Ling[1], Zihao Zhang[1], Yiming Wang[1], Yilin Wang[1], Xijian Zhang[1], Pingrang Hua[2], Qingpu Wang[1], Aimin Song[1,3][\*], Yifei Zhang[1][\*\*]*

[1]Shandong Technology Center of Nanodevices and Integration, School of Microelectronics, Shandong University, Jinan, 250100, China

[2]Department of Opto-electronics and Information Engineering, School of Precision Instruments and Opto-electronics Engineering, Tianjin University, Tianjin, 300072, China

[3]School of Electrical and Electronic Engineering, University of Manchester, Manchester, M13 9PL, United Kingdom

## Abstract

Plasma frequency is the spectral boundary for low-loss propagation and evanescent decay of surface plasmon polariton (SPP) waves, which corresponds to a high cut-off phenomenon and is typically utilized for identifying SPPs. At terahertz (THz) frequencies, a metal line with periodic metallic grooves can mimic the conventional optical SPPs, which is referred to as designer SPPs. Theoretically, the plasma frequency of THz SPPs decreases as the groove depth increases. Here, by replacing the metallic grooves with graphene sheets, dynamically sweeping SPP plasma frequency is demonstrated for the first time. The metal-graphene hybrid structure comprises a metal line with periodic graphene grooves, a thin-layer ion gel for gating graphene, and

[\*] Corresponding author.
[\*\*] Corresponding author.
E-mail: A.Song@manchester.ac.uk (Aimin Song), yifeizhang@sdu.edu.cn (Yifei Zhang)
[a] Authors contribute equally to this work.

metallic tips for uniforming gate field. As the chemical potential changes, the average conductivity of graphene is modulated so that the effective depth of the graphene grooves changes, which sweeps the plasma frequency of THz SPPs consequently. Both simulated and experimental data demonstrate a red shift of plasma frequency from 195 to 180 GHz at a low bias from -0.5 to 0.5 V. The proposed structure reveals a novel approach to control the on/off status of SPP propagation in the THz range.

# 1. Introduction

Surface plasmon polaritons (SPPs) are electromagnetic surface waves propagating along the interface of metal and dielectric with opposite dielectric constants, which exhibit fabulous features of non-diffraction limit and strong electric field confinement.[1,2] They have found many attractive applications at optical, terahertz (THz) and microwave frequencies,[3,4] such as highly sensitive biosensing,[5,6] super-resolution imaging,[7,8] ultrahigh-efficiency photovoltaic,[9] etc. In physical terms, SPPs are dispersive slow-waves, whose dispersive curves gradually deviate from the light line and approach a plasma frequency as the frequency increases.[2,4] This plasma frequency is a boundary for wave propagation and evanescent decay in the spectrum, which hints SPPs can hardly propagate above the plasma frequency.[10,11] Originally, SPPs were reported in the visible and infrared regime, where the noble metals with plasmonic properties show negative permittivity.[12,13] At THz frequencies, where the metals are perfect electric conductors, structured metals with sub-wavelength periodic units can mimic the optical SPPs with similar non-diffraction limit and local field enhancement.[14,15] These THz SPPs address low propagation loss, weak mutual coupling, and simple integration with traditional transmission lines,[13,16] and thus have been utilized in low-profile splitters,[17] filters,[18] antennas,[19] amplifiers,[20] etc. In contrast to the canonical optical SPPs, the dispersion properties of THz SPPs are determined by the dimensions

of the periodic structures, which reveals the propagation of THz SPPs can be tailored.[10, 15, 21]

As passive devices, the metallic SPP structures have fixed optical properties after design. By combining with active stimuli, their fixed properties can be dynamically modulated to achieve versatile functions.[22] Recently, several approaches have been proposed to modulate SPPs actively, such as ferroelectric materials,[23] temperature-sensitive materials,[24] varactor diodes,[25] Schottky diodes,[26] and 2-D materials.[27] Among these approaches, graphene is a promising candidate due to its unique properties, such as high electron mobility, good optical transparency, and excellent thermal conductivity.[28, 29] The electrical conductivity of graphene can be modulated within a large range over one order with electric field,[30] optical pump,[31] and chemical diffusion.[32] To date, active SPP devices controlled with graphene have been reported for frequency modulation in narrow-band resonators and amplitude modulation in broadband structures. For the former, periodic graphene strips[33] and metal strips on top of a continuous graphene layer[34] provide sharp resonance with tunable frequency; for the latter, placing graphene on a metal line with periodic grooves can actively modulate the SPP attenuation within a broad bandwidth.[35, 36] However, to the best of the authors' knowledge, active sweep of plasma frequency has not been reported with graphene and other stimuli yet, which may provide a new function for switching SPPs between low-loss propagation and large decay continuously.

In this paper, we propose sweeping the plasma frequency of THz SPPs in a metal-graphene hybrid structure for the first time. In the proposed structure, the periodic metallic grooves on a metal line are substituted with graphene grooves, and a 70-um thick solution-processed poly (styrenesulfonic acid sodium salt) (PSSNa) was spin-coated on the graphene to provide gate bias. By sweeping the bias voltage, the average conductivity of graphene is modulated continuously, which correspondingly changes the effective depth of the graphene grooves and, thus, the plasma frequency of SPPs. To provide relatively uniform gate field for graphene, metallic tips are designed on the metal line, which slightly changes the initial plasma frequency. The active modulation of plasma frequency is first demonstrated with finite-element-method simulation, and then experimentally verified with a fabricated device. The tuning range is from 195 to 180 GHz with a low bias voltage of less than 1 V.

## 2. Methods and experimental details

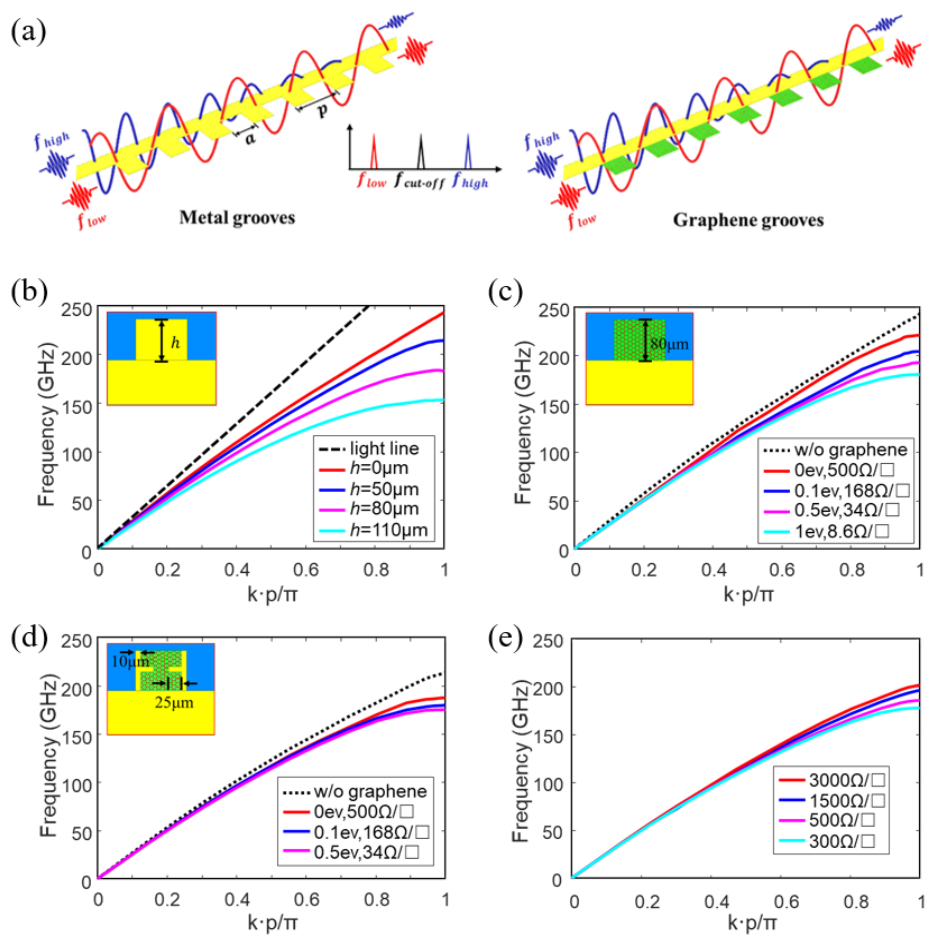

**Figure 1.** THz SPPs and their simulated dispersion curves. (a) Metal and metal-graphene hybrid structures for THz SPPs. (b) Dispersion curves for metal grooves ($p$=0.21 mm, $a$=0.1 mm, $h$=0, 50, 80, and 110 μm). The dashed line is for the light in free space. (c) Dispersion curves for graphene grooves at various chemical potentials ($h$= 80 μm). Dispersion curves for graphene grooves and bias tips at (d) various chemical potentials and (e) various sheet resistances.

*2.1 Eigen mode analysis*

Typically, a metal line with periodic grooves guides THz SPPs, as shown in Figure 1(a), whose propagation is dispersive as the classic optical SPPs. The propagation constant of the THz SPPs is given by[37]

$$\beta = \sqrt{\varepsilon_{eff} k_0^2 + (a/p)^2 \varepsilon_{eff} k_0^2 tan^2 \left(k_0 \sqrt{\varepsilon_{eff}} h\right)}, \quad (1)$$

where $a$ is the width of the groove, $p$ is the lattice period, $h$ is the groove depth, $k_0$ is the wave number of the light in free space, and $\varepsilon_{eff}$ is the effective permittivity for metal grooves. The corresponding plasma frequency is approximately estimated as[37]

$$\omega_p = \pi c_0 / 2\sqrt{\varepsilon_{eff}} h, \quad (2)$$

where $c_0$ represents the free space light speed. Note that the plasma frequency decreases as the groove depth $h$ increases.

To precisely predict the plasma frequency of THz SPPs, the dispersion curves are simulated using Eigen Mode Solver in Ansys High Frequency Structural Simulator (HFSS). The inset of Figure 1(b) depicts the conventional metal unit on a silicon substrate for SPP mode analysis, whose dispersion curves are illustrated in Figure 1(b). For the fixed groove width ($a$=0.1 mm) and lattice period ($p$=0.21 mm), the plasma frequency decreases as the groove depth $h$ increases, which is consistent with Equations (1) and (2). In addition, the SPPs show slow-wave effect as their dispersion curves are located on the right side of the light line. The plasma frequency of the metal SPPs is 180 GHz at $h$=80 μm. Considering the spectral range of our measurement system, $h$=80 μm is chosen for the following analysis.

To actively change the groove depth, the metal grooves are substituted with graphene grooves, as shown in the inset of Figure 1(c). At THz frequencies, the conductivity of graphene ($\sigma_g$) is typically determined by the intraband term of the Kubo formula.[28, 29] The effective depth of graphene grooves can be estimated using the attenuation length ζ, which is the distance where the electromagnetic waves attenuate to $1/e$.[37] At terahertz frequencies,

$$\zeta = -\frac{|\sigma_g|^2}{2\omega\varepsilon_d\sigma_g''}, \qquad \sigma_g'' < 0, \tag{3}$$

where $\omega$ is the angular frequency, $\varepsilon_d$ is the relative permittivity of the dielectric, and $\sigma_g''$ is the imaginary part of $\sigma_g$.[37] Notably, the effective groove depth enlarges as the graphene conductivity increases. The simulated dispersion curves of graphene grooves are illustrated in Figure 1(c). As the graphene conductivity increases, the plasma frequency continuously red shifts from 220 to 180 GHz.

However, it is challenging to apply uniform bias field to the whole graphene grooves as the bias electrode is asymmetric in Figure 1(c). To get more uniform field, metallic bias tips are designed on the metal line, which slightly shifts the initial plasma frequency, as shown in Figure 1(d). As the chemical potential increases from 0 to 0.5 eV, the plasma frequency sweeps from 185 to 175 GHz. Here, it should be noted that the transferred large-area chemical vapor deposition (CVD) graphene typically has larger average resistivity than that calculated from the Kubo formula.[39] Therefore, the actual devices may have different modulation range, as illustrated in Figure 1(e). The sheet resistance extracted from the direct-current (DC) test of a graphene field effect transistor is approximately from 3000 to 300 Ω/□. Correspondingly, the simulated plasma frequency is from 200 to 180 GHz.

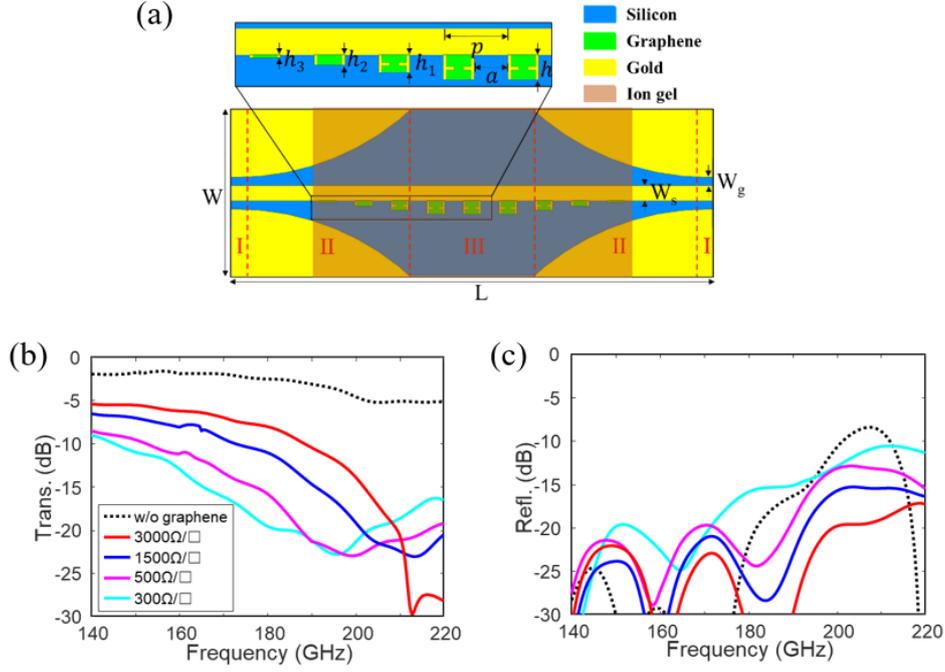

**Figure 2.** (a) HFSS 3-D model of the metal-graphene hybrid structure with CPW feeding ($h_1$=58 μm, $h_2$=34 μm, $h_3$=10 μm, W=1.5 mm, L=2.8 mm, $W_s$=87 μm, $W_g$=5 μm). Simulated (b) transmittance and (c) reflectance of the proposed structure at various sheet resistances.

*2.2 Full wave simulation*

Based on the above eigen mode analysis, a metal-graphene hybrid model with bias tips and coplanar waveguide (CPW) feeding is designed and simulated using Driven Mode Solver, as shown in Figure 2(a). It consists of two 50-Ω feeding CPWs (I), two transitions for transferring CPW mode to SPP mode (II), and one hybrid SPP waveguide with periodic graphene grooves and bias tips (III) on a 200-μm thick silicon substrate. Their dimensions are listed in the caption of Figure 2, whose design principles are the same as the metal SPP waveguide in our previous paper.[26] Here, graphene is modeled as a resistive sheet with zero thickness, and its sheet resistance varies from 3000 to 300 Ω/□. The details of the sheet resistance can be found in the following section. A 70-μm thick ion gel layer is designed on the surface, whose relative permittivity is around 3, as reported in Ref.[40, 41] The simulated transmittance and reflectance are shown in Figures 2(b) and 2(c), respectively. For the metal structure without graphene, the plasma frequency is around 210 GHz. With the presence of graphene, the plasma frequency can be modulated from 180 to 200 GHz, showing reasonable consistency with the eigen mode simulation in Figure 1(e). To better understand the mechanism, the electric field distributions of the proposed devices with various sheet resistance for graphene at 180 GHz are shown in Figure 3. As the graphene conductivity increases, the field concentration moves from the middle of the graphene squares, as depicted in Figure 3(a), to the outer edge, as depicted in Figure 3(d), which reveals the effective depth of the graphene grooves increases. Consequently, the plasma frequency red shifts, and the attenuation at 180 GHz enlarges significantly due to the evanescent decay, see the magnitude of the output signals in Figure 3.

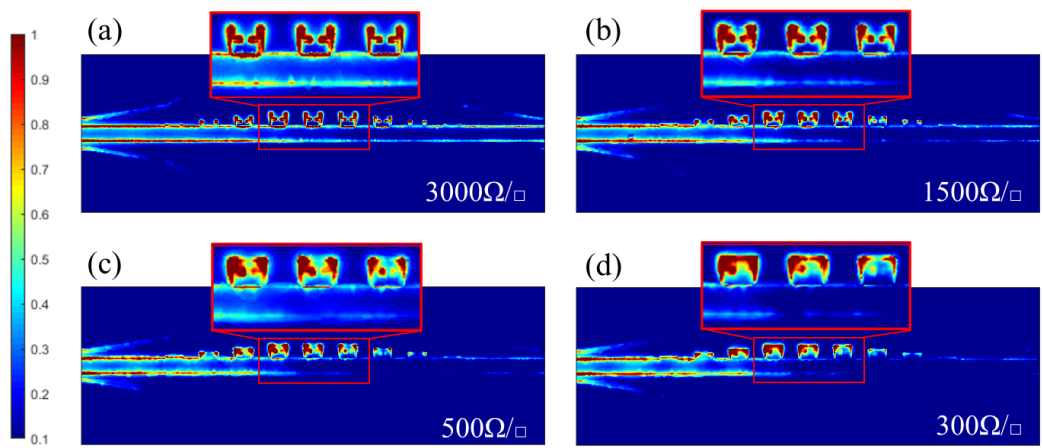

**Figure 3.** Electric field distribution of the hybrid SPP structure with various sheet resistance for graphene. (a) 3000 Ω/□, (b) 1500 Ω/□, (c) 500 Ω/□, and (d) 300 Ω/□. The concentrated fields move towards the outer edge of graphene as the resistance decreases.

*2.3 Fabrication and measurement*

The proposed structure was fabricated following the steps depicted in Figure 4(a).

Firstly, metal electrodes were fabricated on a 200-μm thick p-type silicon substrate with a 100-nm SiO$_2$ layer using standard ultraviolet photolithography, e-beam deposition, and lift-off process.[26] Next, a CVD graphene was wet transferred onto the substrate,[42] and patterned with photolithography and dry etching. Finally, a thin-layer of PSSNa was spin-coated onto the device as electrolyte to form an electric double-layer capacitor (EDLC) for gating graphene. The fabricated device is illustrated in Figure 4(b), and the scanning electron microscope (SEM) image of the PSSNa layer is illustrated in Figure 4(c). It can be seen that the thickness of PSSNa is 70 μm.

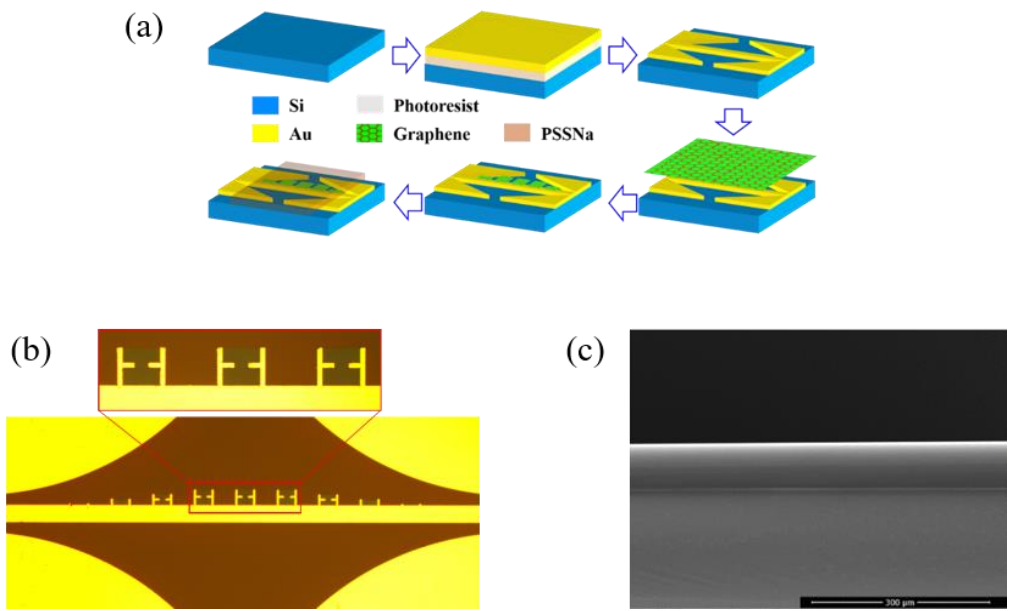

**Figure 4.** Fabricated metal-graphene hybrid SPP structure with bias tips. (a) Fabrication steps, (b) image of the fabricated structure, (c) SEM image of the spin-coated ion gel.

For measurement, two 50-Ω ground-signal-ground (G-S-G) probes integrated with Agilent programmable network analyzer (PNA) were employed to characterize the transmittance and reflectance of the fabricated structure, as shown in Figure 5(a). The probes were first calibrated with short-load-open-through (SLOT) method and then launched on the CPWs of the device for feeding and receiving signals. During the modulation test, the DC-bias is applied on the EDLC between the CPW ground and signal electrodes for sweeping the graphene conductivity, and the spectral variation is characterized by using PNA.

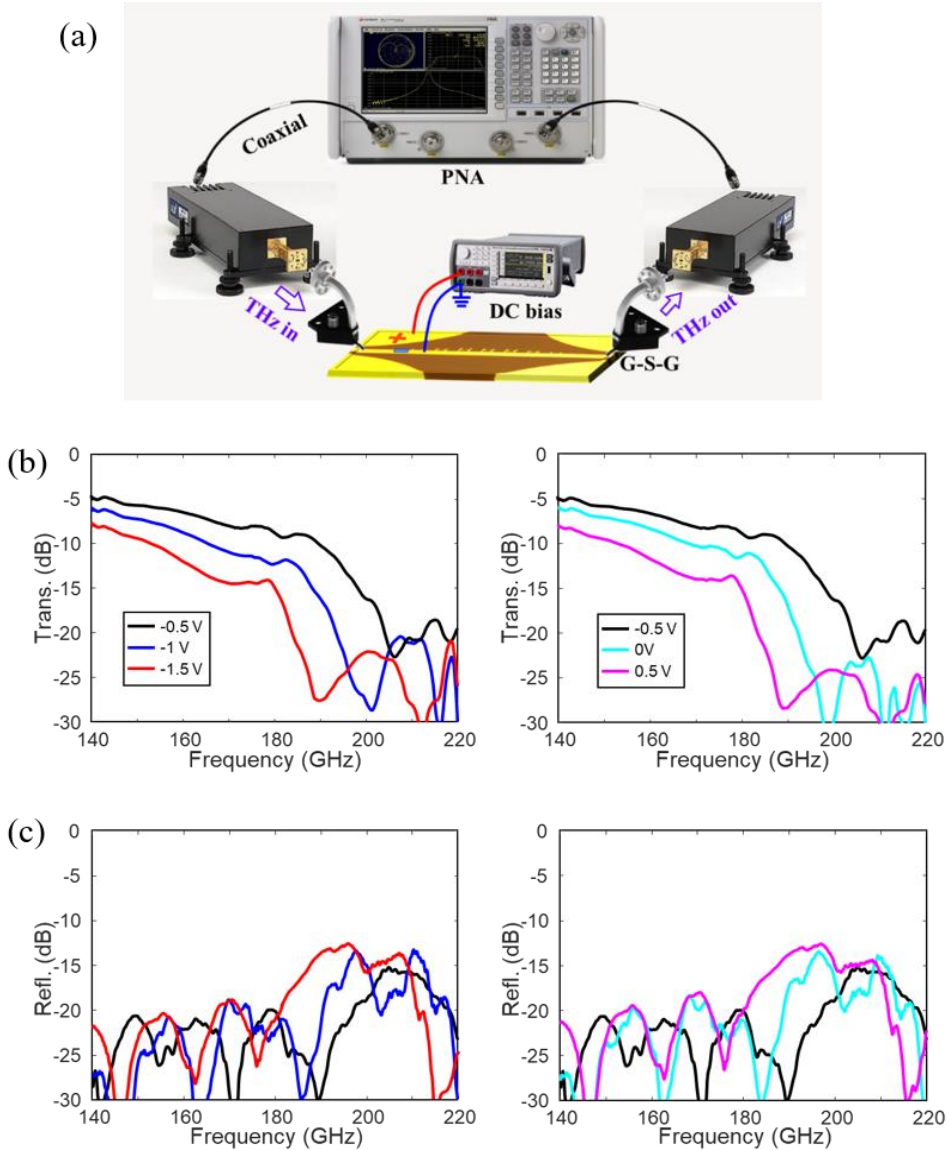

**Figure 5.** (a) Measurement system setup and measured (b) transmittance and (c) reflectance. The plasma frequency blue shifts as the bias changes from -1.5 to -0.5 V, and red shifts as the bias further changes from -0.5 to 0.5 V.

## 3. Results and discussion

*3.1 Tunable plasma frequency*

The characterized transmittance and reflectance at various bias voltages are shown in Figures 5(b) and 5(c), showing clear cut-off response at high frequencies. The cut-off frequency is typically referred to as the plasma frequency of THz SPPs, which is continuously modulated with various bias voltages. As the bias sweeps from -1.5 to -0.5 V, the plasma frequency blue shifts from 181 to 195 GHz due to the shrinking effective groove depth; as the bias further sweeps from -0.5 to 0.5 V, the plasma frequency red shifts from 195 to 180 GHz due to the increasing effective groove depth. Consequently, the transmittance is modulated from -10 dB to -28 dB at 190 GHz due to the plasma frequency sweeping, showing a good on/off ratio of 63 and matching well with the simulated field distribution in Figure 3. Meanwhile, the transmittance under the plasma frequency is also modulated due to the graphene absorption, while the reflectance is relatively stable. The amplitude modulation range of the transmittance is 3 and 6 dB at 140 and 170 GHz, respectively, and the corresponding on/off ratio is 2 and 25, respectively. As the frequency approaches the plasma frequency, the enhanced slow-wave effect enlarges the graphene absorption. Here, it should be noted that our structure achieves plasma frequency modulation and amplitude modulation simultaneously in different spectral ranges in subwavelength scales.

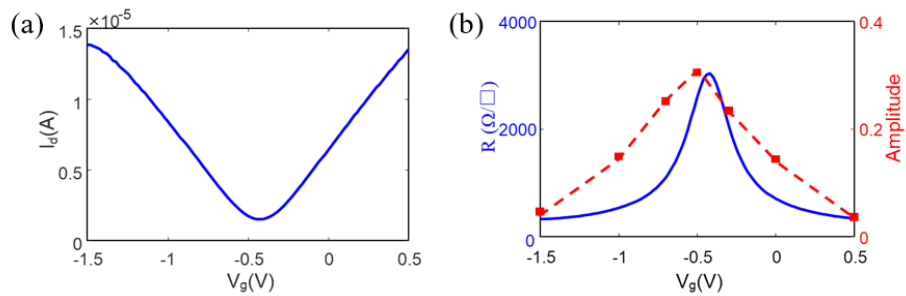

**Figure 6.** (a) I-V response of a graphene transistor using PSSNa as the top gate. (b) The amplitude of transmission at 190 GHz as a function of gate bias.

*3.2 Comparison with DC data*

As can be seen in Figure 5, the modulation trend of the plasma frequency predicts that the Dirac point of the graphene used here is around -0.5 V. To verify this, a graphene transistor fabricated on the same substrate was characterized, whose current-voltage (I-V) curve is shown in Figure 6(a). The DC curve reveals that the Dirac point is at -0.43 V, showing good consistency with the spectral data. The on/off ratio for the current is 9.4, which is among the best of the reported works.[27,35,36] The channel length and width of the transistor is 300 and 350 μm, respectively. Thus, the calculated sheet resistance for graphene sweeps approximately from 3000 to 300 Ω/□, as shown in Figure 6(b). Also, Figure 6(b) illustrates the transmission amplitude at 190 GHz as a function of the applied bias. It can be found that the transmission is modulated from 0.31 to 0.04, as the bias voltage sweeps from -1.5 to 0.5 V, which can be mainly attributed to the change of the plasma frequency. As the graphene resistance increases, the effective groove depth becomes smaller, and the transmission enhances, which implies a blue shift of the plasma frequency; as the resistance decreases, the effective groove depth enlarges, which corresponds to a red shift of the plasma frequency as well as the transmission decrease. Here, it is worthy to mention that dynamic control of plasma frequency offers an unusual mechanism of controlling evanescent decay for novel active attenuators and filters. In addition, the applied sub-volt bias in our approach is smaller than the other reported works,[25-27,35,36,43] which is fully compatible with complementary metal–oxide–semiconductor (CMOS) technologies.

## 4. Conclusions

In summary, actively sweeping plasma frequency of THz SPPs is demonstrated by substituting metal grooves with graphene grooves on a metal line. With a thin-layer of PSSNa gating the graphene grooves, the effective groove depth can be modulated by sweeping the graphene chemical potential, which shifts the plasma frequency in bidirectional trend. The sweeping range in the spectrum is from 180 to 195 GHz with a low bias from -0.5 V to 0.5 V. This work provides a new method for actively modulating THz SPPs from propagation to evanescent attenuation at CMOS-compatible voltages.

**Acknowledgments**

The authors would like to thank the following grants for supporting this work: National Key Research and Development Program of China (2016YFA0301200); Key Research and Development Program of Shandong Province (2019JZZY020109); National Natural Science Foundation of China (61701283); Engineering and Physical Sciences Research Council (EP/N021258/1); China Postdoctoral Science Foundation funded project (2018T110689, 2017M622201); Postdoctoral Innovation Program of Shandong Province (20171006).

**References:**

1. Barnes, W. L.; Dereux, A.; Ebbesen, T. W., Surface Plasmon Subwavelength Optics. *Nature* **2003,** *424*, 824-830. https://doi.org/10.1038/nature01937

2. Hibbins, A. P.; Evans, B. R.; Sambles, J. R., Experimental Verification of Designer Surface Plasmons. *Science* **2005,** *308*, 670-672. https://doi.org/10.1126/science.1109043

3. Zhang, X.; Xu, Q.; Xia, L.; Li, Y.; Gu, J.; Tian, Z.; et al., Terahertz Surface Plasmonic Waves: a Review. *Advanced Photonics* **2020,** *2*, 1. https://doi.org/10.1117/1.AP.2.1.014001

4. Berini, P.; De Leon, I., Surface Plasmon Polariton Amplifiers and Lasers. *Nat. Photonics* **2012,** *6*, 16-24. https://doi.org/10.1038/nphoton.2011.285

5. Anker, J. N.; Hall, W. P.; Lyandres, O.; Shah, N. C.; Zhao, J.; Van Duyne, R. P., Biosensing with Plasmonic Nanosensors. *Nat. Mater.* **2008,** *7*, 442-453.

https://doi.org/10.1038/nmat2162

6. Chen, X.; Fan, W., Ultrasensitive Terahertz Metamaterial Sensor Based on Spoof Surface Plasmon. *Sci. Rep.* **2017,** *7*, 2092. https://doi.org/10.1038/s41598-017-01781-6

7. Jones, A. C.; Olmon, R. L.; Skrabalak, S. E.; Wiley, B. J.; Xia, Y. N.; Raschke, M. B., Mid-IR Plasmonics: Near-Field Imaging of Coherent Plasmon Modes of Silver Nanowires. *Nano Lett.* **2009,** *9*, 2553-2558. https://doi.org/10.1021/nl900638p

8. Tang, H. H.; Liu, P. K., Terahertz Far-Field Superresolution Imaging Through Spoof Surface Plasmons Illumination. *Opt. Lett.* **2015,** *40*, 5822-5. https://doi.org/10.1364/ol.40.005822

9. Polman, A.; Atwater, H. A., Photonic Design Principles for Ultrahigh-Efficiency Photovoltaics. *Nat. Mater.* **2012,** *11*, 174-177. https://doi.org/10.1038/nmat3263

10. Pendry, J. B.; Martín-Moreno, L.; Garcia-Vidal, F. J., Mimicking Surface Plasmons with Structured Surfaces. *Science* **2004,** *5685*, 847-848. https://doi.org/10.1126/science.1098999

11. Shin, H.; Fan, S., All-angle Negative Refraction for Surface Plasmon Waves Using a Netal-Dielectric-Metal Structure. *Phys. Rev. Lett.* **2006,** *96*, 073907. https://doi.org/10.1103/PhysRevLett.96.073907

12. Gan, Q.; Fu, Z.; Ding, Y. J.; Bartoli, F. J., Ultrawide-Bandwidth Slow-Light System Based on THz Plasmonic Graded Metallic Grating Structures. *Phys. Rev. Lett.* **2008,** *100*, 256803. https://doi.org/10.1103/PhysRevLett.100.256803

13. Ma, H. F.; Shen, X. P.; Cheng, Q.; Jiang, W. X.; Cui, T. J., Broadband and High-Efficiency Conversion from Guided Waves to Spoof Surface Plasmon Polaritons. *Laser*

*Photonics Rev.* **2014,** *8*, 146-151. https://doi.org/10.1002/lpor.201300118

14. Zhang, H. C.; Cui, T. J.; Zhang, Q.; Fan, Y. F.; Fu, X. J., Breaking the Challenge of Signal Integrity Using Time-Domain Spoof Surface Plasmon Polaritons. *ACS Photonics* **2015,** *2*, 1333-1340. https://doi.org/ 10.1021/acsphotonics.5b00316

15. Garcia-Vidal, F. J.; Martín-Moreno, L.; Pendry, J. B., Surfaces with Holes in Them: New Plasmonic Metamaterials. *J. Opt.* **2005,** *7*, S97-S101. https://doi.org/10.1088/1464-4258/7/2/013

16. Zhang, H. C.; Zhang, Q.; Liu, J. F.; Tang, W.; Fan, Y.; Cui, T. J., Smaller-Loss Planar SPP Transmission Line than Conventional Microstrip in Microwave Frequencies. *Sci. Rep.* **2016,** *6*, 23396. https://doi.org/10.1038/srep23396

17. Xi, G.; Jin, H. S.; Shen, X.; Hui, F. M.; Wei, X. J.; Li, L.; et al., Ultrathin Dual-Band Surface Plasmonic Polariton Waveguide and Frequency Splitter in Microwave Frequencies. *Appl. Phys. Lett.* **2013,** *102*, 824. https://doi.org/10.1063/1.4802739

18. Yin, J. Y.; Ren, J.; Zhang, H. C.; Pan, B. C.; Cui, T. J., Broadband Frequency-Selective Spoof Surface Plasmon Polaritons on Ultrathin Metallic Structure. *Sci. Rep.* **2015,** *5*, 8165. https://doi.org/10.1038/srep08165

19. Xu, J. J.; Hao, C. Z.; Qian, Z.; Cui, T. J., Efficient Conversion of Surface-Plasmon-Like Modes to Spatial Radiated Modes. *Appl. Phys. Lett.* **2015,** *106*, 10681. https://doi.org/10.1063/1.4905580

20. Zhang, H. C.; Liu, S.; Shen, X. P.; Chen, L. H.; Li, L. M.; Cui, T. J., Broadband Amplification of Spoof Surface Plasmon Polaritons at Microwave Frequencies. *Laser Photonics Rev.* **2015,** *9*, 83-90. https://doi.org/10.1002/lpor.201400131

21. Shen, X. P.; Cui, T. J., Planar Plasmonic Metamaterial on a Thin Film with Nearly

Zero Thickness. *Appl. Phys. Lett.* **2013**, *102*, 211909. https://doi.org/10.1063/1.4808350

22. Jie, X.; Hao, C. Z.; Tang, W.; Jian, G.; Cheng, Q.; Li, W., Transmission-Spectrum-Controllable Spoof Surface Plasmon Polaritons Using Tunable Metamaterial Particles. *Appl. Phys. Lett.* **2016,** *108*, 824-848. https://doi.org/10.1063/1.4950701

23. Luo, J.; He, J.; Apriyana, A.; Feng, G.; Huang, Q.; Zhang, Y. P., Tunable Surface-Plasmon-Polariton Filter Constructed by Corrugated Metallic Line and High Permittivity Material. *IEEE Access* **2018,** *6*, 10358-10364. https://doi.org/10.1109/ACCESS.2018.2800158

24. Zhang, T.; Zhang, Y. X.; Shi, Q. W.; Yang, X. B.; Liang, S. X.; Fang, Y.; et al., On-Chip THz Dynamic Manipulation Based on Tunable Spoof Surface Plasmon Polaritons. *IEEE Electron Device Lett.* **2019,** *40*, 1844-1847. https://doi.org/10.1109/LED.2019.2940144

25. Zhang, X.; Tang, W.; Zhang, H.; Xu, J.; Bai, G. D.; Liu, J.; et al., A Spoof Surface Plasmon Transmission Line Loaded with Varactors and Short-Circuit Stubs and Its Application in Wilkinson Power Dividers. *Adv. Mater. Technol.* **2018,** *3*, 1800046. https://doi.org/10.1002/admt.201800046

26. Zhang, Y. F.; Ling, H. T.; Chen, P. J.; Qian, P. F.; Shi, Y. P.; Wang, Y. M.; et al., Tunable Surface Plasmon Polaritons with Monolithic Schottky Diodes. *ACS Appl. Electron Mat.* **2019,** *1*, 2124-2129. https://doi.org/10.1021/acsaelm.9b00499

27. Tian, D.; Kianinejad, A.; Zhang, A. X.; Chen, Z. N., Graphene-Based Dynamically Tunable Attenuator on Spoof Surface Plasmon Polaritons Waveguide. *IEEE Microw. Wireless Compon. Lett.* **2019,** *29*, 388-390. https://doi.org/10.1109/LMWC.2019.2913964

28. Islam, M. S.; Sultana, J.; Biabanifard, M.; Vafapour, Z.; Nine, M. J.; Dinovitser, A.; et al., Tunable Localized Surface Plasmon Graphene Metasurface for Multiband Superabsorption and Terahertz Sensing. *Carbon* **2020,** *158*, 559-567. https://doi.org/10.1016/j.carbon.2019.11.026

29. Li, J. T.; Li, J.; Zheng, C. L.; Wang, S. L.; Li, M. Y.; Zhao, H. L.; et al., Dynamic Control of Reflective Chiral Terahertz Metasurface with a New Application Developing in Full Grayscale Near Field Imaging. *Carbon* **2021,** *172*, 189-199. https://doi.org/10.1016/j.carbon.2020.09.090

30. Wu, Y.; La-o-vorakiat, C.; Qiu, X.; Liu, J.; Deorani, P.; Banerjee, K.; et al., Graphene Terahertz Modulators by Ionic Liquid Gating. *Adv. Mater.* **2015,** *27*, 1874-9. https://doi.org/10.1002/adma.201405251

31. Tasolamprou, A. C.; Koulouklidis, A. D.; Daskalaki, C.; Mavidis, C. P.; Kenanakis, G.; Deligeorgis, G.; et al., Experimental Demonstration of Ultrafast THz Modulation in a Graphene-Based Thin Film Absorber Through Negative Photoinduced Conductivity. *ACS Photonics* **2019,** *6*, 720-727. https://doi.org/10.1021/acsphotonics.8b01595

32. Samuels, A. J.; Carey, J. D., Engineering Graphene Conductivity for Flexible and High-Frequency Applications. *ACS Appl. Mater. Interfaces* **2015,** *7*, 22246-55. https://doi.org/10.1021/acsami.5b05140

33. Ju, L.; Geng, B.; Horng, J.; Girit, C.; Martin, M.; Hao, Z.; et al., Graphene Plasmonics for Tunable Terahertz Metamaterials. *Nat. Nanotechnol.* **2011**, *6*, 630-634. https://doi.org/10.1038/nnano.2011.146

34. Jadidi, M. M.; Sushkov, A. B.; Myers-Ward, R. L.; Boyd, A. K.; Daniels, K. M.; Gaskill, D. K.; et al., Tunable Terahertz Hybrid Metal-Graphene Plasmons. *Nano Lett.* **2015**, *15*, 7099-7104. https://doi.org/10.1021/acs.nanolett.5b03191

35. Zhang, A. Q.; Lu, W. B.; Liu, Z. G.; Wu, B.; Chen, H., Flexible and Dynamically Tunable Attenuator Based on Spoof Surface Plasmon Polaritons Waveguide Loaded with Graphene. *IEEE Trans. Antennas Propag.* **2019**, *67*, 5582-5589. https://doi.org/10.7567/1882-0786/ab25fb

36. Chen, Z. P.; Lu, W. B.; Liu, Z. G.; Zhang, A. Q.; Wu, B.; Chen, H., Dynamically Tunable Integrated Device for Attenuation, Amplification, and Transmission of SSPP Using Graphene. *IEEE Trans. Antennas Propag.* **2020,** *68*, 3953-3962. https://doi.org/10.1109/TAP.2020.2963946

37. Liu, X.; Feng, Y.; Zhu, B.; Zhao, J.; Jiang, T., High-Order Modes of Spoof Surface Plasmonic Wave Transmission on Thin Metal Film Structure. *Opt. Express* **2013,** *21*, 31155-65. https://doi.org/10.1364/OE.21.031155

38. Hanson, G. W., Dyadic Green's Functions and Guided Surface Waves for a Surface Conductivity Model of Graphene. *J. Phys. D* **2008,** *103*, 064302. https://doi.org/10.1063/1.2891452

39. Kholmanov, I. N.; Magnuson, C. W.; Aliev, A. E.; Li, H. F.; Zhang, B.; Suk, J. W.; et al., Improved Electrical Conductivity of Graphene Films Integrated with Metal Nanowires. *Nano Lett.* **2012,** *12*, 5679-5683. https://doi.org/10.1021/nl302870x

40. Yamamoto, K.; Tani, M.; Hangyo, M., Terahertz Time-Domain Spectroscopy of Imidazolium Ionic Liquids. *J. Phys. Chem. B* **2007,** *111*, 4854-4859. https://doi.org/10.1021/jp067171w

41. Ju, L.; Geng, B.; Horng, J.; Girit, C.; Martin, M.; Hao, Z.; Bechtel, H. A.; et al., Graphene Plasmonics for Tunable Terahertz Metamaterials. *Nat. Nanotechnol.* **2011,** *6*, 630-4. https://doi.org/10.1038/nnano.2011.146

42. Liu, J.; Qian, Q.; Zou, Y.; Li, G.; Jin, Y.; Jiang, K.; et al., Enhanced Performance

of Graphene Transistor with Ion-Gel Top Gate. *Carbon* **2014,** *68*, 480-486. https://doi.org/10.1016/j.carbon.2013.11.024

43. Yi, Y.; Zhang, A. Q., A Tunable Graphene Filtering Attenuator Based on Effective Spoof Surface Plasmon Polariton Waveguide. *IEEE Trans. Microw. Theory* **2020**, *68*, 5169-5177. https://doi.org/10.1109/Tmtt.2020.3026694